\documentclass{article}
\usepackage[T1]{fontenc}
\usepackage[utf8]{inputenc}

\usepackage{dcase2024,amsmath,graphicx,times,booktabs, tabularx,bm,amssymb, amsfonts}
\usepackage{comment}
\usepackage[hyphens]{url}
\usepackage{hyperref}
\usepackage{multirow}
\usepackage{xspace}
\usepackage{xcolor}

\newcommand{\dcaseTwentyFiveTask}{DCASE25T4\xspace}
\newcommand{\dcaseTwentySixSTask}{DCASE26T4\xspace}
\newcommand{\newmodule}{\texttt{SpAudSyn}\xspace}
\renewcommand{\thefootnote}{\fnsymbol{footnote}}

\title{Description and Discussion on DCASE 2026 Challenge Task 4:\\ Spatial Semantic Segmentation of Sound Scenes
}

\makeatletter
\def\thirdlinename#1{\gdef\@thirdlinename{{\em #1}\\}}
\gdef\@thirdlinename{}

\renewcommand{\@maketitle}{\newpage
  \null
  \vskip 1em
  \begin{center}
    {\large\bf \@title \par}
    \vskip 1.5em
    {\large\lineskip .5em
     \begin{tabular}[t]{c}
       \@name
       \ifx\@secondlinename\@empty\else \\[-1em]\@secondlinename\fi
       \ifx\@thirdlinename \@empty\else \\[-1em]\@thirdlinename \fi
       \\ \@address
     \end{tabular}\par}
  \end{center}
  \vskip 1.5em}
\makeatother

\name{Binh Thien Nguyen$^{*1}$,
      Masahiro Yasuda$^{*1}$,
      Noboru Harada$^{1}$,
      Romain Serizel$^{2}$,
      Mayank Mishra$^{2}$}

\secondlinename{Marc Delcroix$^{1}$,
                Carlos Hernandez-Olivan$^{1}$,
                Shoko Araki$^{1}$}

\thirdlinename{Daiki Takeuchi$^{1}$,
               Tomohiro Nakatani$^{1}$,
               Nobutaka Ono$^{3}$}

\address{$^1$ NTT, Inc., Japan, masahiro.yasuda@ntt.com\\          
        $^2$  University de Lorraine, CNRS, Inria, Loria, France\\
        $^3$  Tokyo Metropolitan University, Japan\\}

\begin{document}

\ninept
\maketitle

\vspace{-10pt}

\begin{abstract}
This paper presents an overview of the Detection and Classification of Acoustic Scenes and Events (DCASE) 2026 Challenge Task 4, Spatial Semantic Segmentation of Sound Scenes (S5).
The S5 task focuses on the joint detection and separation of sound events in complex spatial audio mixtures, contributing to the foundation of immersive communication.
First introduced in DCASE 2025, the S5 task continues in DCASE 2026 Task 4 with key changes to better reflect real-world conditions, including allowing mixtures to contain multiple sources of the same class and to contain no target sources.
In this paper, we describe task setting, along with the corresponding updates to the evaluation metrics and dataset.
The experimental results of the submitted systems are also reported and analyzed. 
The official access point for data and code is \url{https://github.com/nttcslab/dcase2026_task4_baseline}.
\end{abstract}


\vspace{-5pt}
\section{Introduction}
\label{sec:intro}

\begingroup
  \renewcommand\thefootnote{}  
  \hypersetup{hidelinks}
  \footnote{This work was partially supported by JST Strategic International Collaborative Research Program (SICORP), Grant Number JPMJSC2306, Japan. This work was partially supported by the Agence Nationale de la Recherche (Project Confluence, grant number ANR-23-EDIA-0003).\\
* These authors contributed equally to this work.}%
  \addtocounter{footnote}{-1} 
\endgroup

Spatial semantic segmentation of sound scenes (S5) refers to the task of identifying and separating individual sound events from complex spatial audio signals.
It takes a multi-channel mixture as input and produces a set of estimated single-channel source signals, each associated with its corresponding sound event class label.
S5 supports the development of technologies across a wide range of applications, including immersive communication, extended reality (XR) systems, and acoustic scene monitoring in smart and assisted living environments.

\begin{figure}[t]
\begin{center}
\includegraphics[width=\columnwidth]{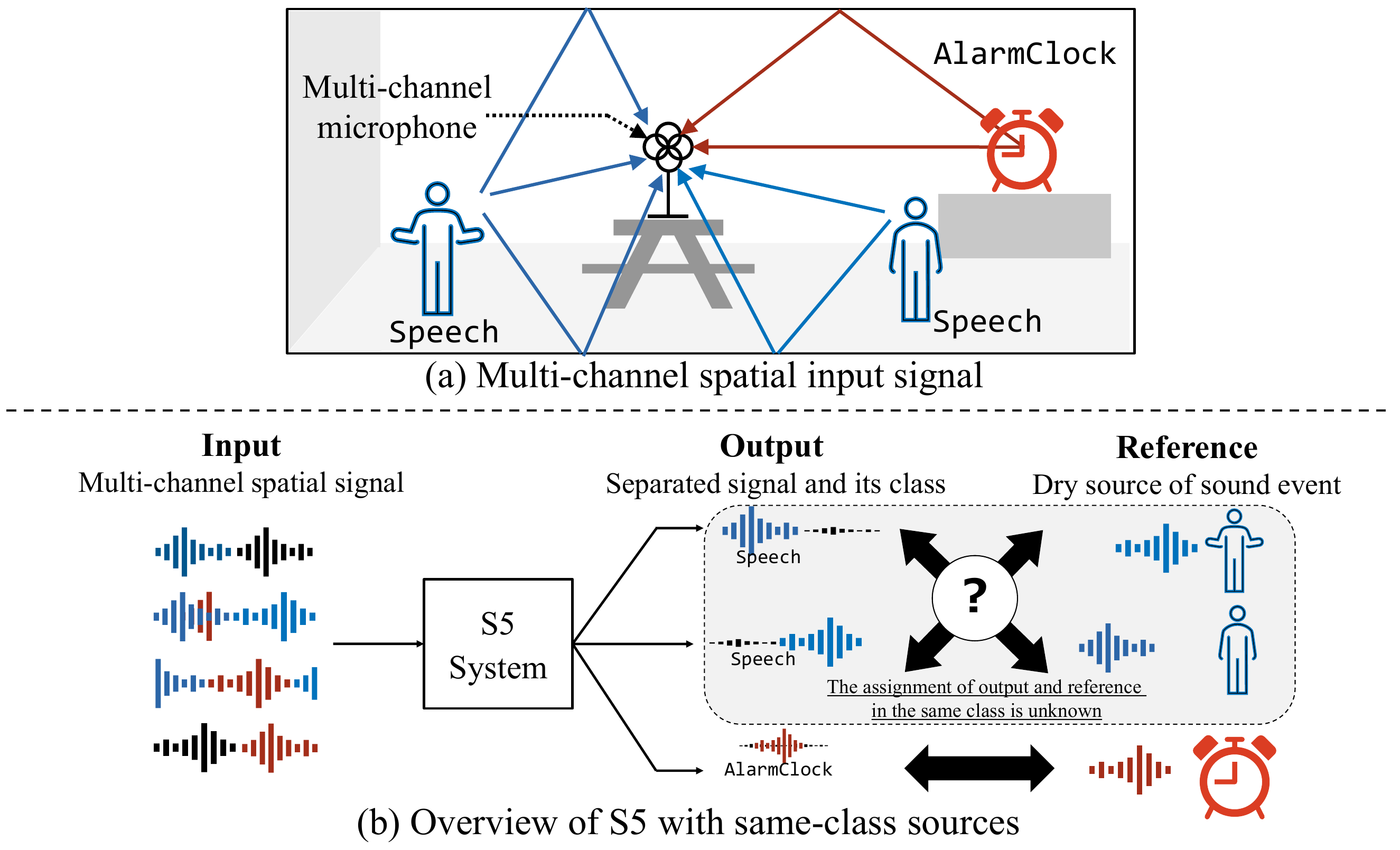}
\end{center}
\vspace{-0.7cm}
\caption{Overview of S5 task with same-class sources.}\label{fig:overview}
\vspace{-0.4cm}
\end{figure}

Detection and Classification of Acoustic Scenes and Events 2025 Challenge Task 4 (DCASE25T4) marks the first challenge to feature the S5 task, specifically focusing on indoor sound event environments recorded using first-order Ambisonics microphone arrays.
Its baseline systems utilize a two-step strategy: first, an audio tagging (AT) model identifies the source classes present in the mixture; then, a label-queried source separation (LQSS) model separates the corresponding audio signals.
The dataset included a variety of target sound classes, with overlapping events of varying durations, along with environmental noise and interference sounds, challenging systems to accurately separate and label simultaneous sources.
To reproduce realistic acoustics, all dataset components used for mixture synthesis, including the room impulse responses, were derived from actual recordings rather than purely simulated signals.
This setup enables systems to better generalize to real-world acoustic scenes.
The challenge received submissions from eight teams, totaling 24 systems, most of which surpassed the baseline and demonstrated notable improvements in both sound event detection and separation.
Several new approaches were explored, including iterative schemes that refine predictions by alternating between label estimation and source separation to enhance overall performance.

For the DCASE 2026 Challenge Task 4 (DCASE26T4), we organized a follow-up S5 task based on the DCASE 2025 setting, introducing two modifications to better reflect realistic and challenging conditions in real-world environments.
First, in contrast to DCASE25T4, where labels in a mixture are mutually exclusive, DCASE26T4 allows repeated labels.
In other words, multiple sources from the same class can appear simultaneously within a single mixture.
Such situations frequently occur in real acoustic environments, for example when multiple people talk simultaneously, as illustrated in Fig.~\ref{fig:overview}.
This increases the difficulty of the task, as the ambiguity introduced by repeated labels poses challenges for both the audio tagging and source separation components of most existing S5 systems.
Second, in DCASE26T4, a mixture may contain zero target sound events, whereas in DCASE25T4 each mixture contains at least one target event.
This setting is important for real-world deployments, where systems must operate continuously while target sound events may occur only occasionally.
The system must therefore not only detect when target sound events occur, but also correctly determine when they are absent.
This becomes particularly challenging in the presence of background noise and interference sound events.

\section{Task setting of S5}
This section presents the task setting, notation, and formulation of DCASE26T4, and highlights the key differences from DCASE25T4 \cite{dcase2025t4}.

\subsection{Formulation and notation}
\label{sec:tasksetting}

The S5 system receives as input $\bm{Y} = [\bm{y}^{(1)},\dots, \bm{y}^{(M)}]^\top \in \mathbb{R}^{M \times T}$, which denotes an $M$-channel time-domain mixture signal with duration $T$.
The mixture at channel $m$ can be synthesized as
\begin{equation}\label{eq:mixture}
    \bm{y}^{(m)}=\sum_{k=1}^{K} \bm{h}^{(m)}_k*\bm{s}_{c_k} +
        \left[ \sum_{j=1}^{J} \bm{h}^{(m)}_j*\bm{s}_{c_j} + \bm{n}^{(m)} \right]_\textit{optional},
\end{equation}
where $\bm{s}_{c_k}$ and $\bm{s}_{c_j}$ denote the target and interfering sound events, respectively; $\bm{h}^{(m)}_k$ and $\bm{h}^{(m)}_j$ represent their corresponding room impulse responses (RIRs); and $\bm{n}^{(m)}$ is the noise signal. $K$ and $J$ are the numbers of target and interfering sound events in the mixture.


The output of the system consists of the labels of the target sound events, $\hat{C} = (\hat{c}_1, \ldots, \hat{c}_{\hat{K}})$, together with their associated separated single-channel waveforms measured at a reference microphone, $\hat{S} = (\hat{\bm{s}}_{\hat{c}_1}, \ldots, \hat{\bm{s}}_{\hat{c}_{\hat{K}}})$, where $\hat{\bm{s}}_{\hat{c}_i} \in \mathbb{R}^{T}$.

While the input and output are similar to DCASE25T4, the key difference is that the labels in $C$ and $\hat{C}$ can be duplicated.
In other words, a mixture can include multiple sources belonging to the same class, each located at different positions in the room. The system is therefore expected to separate sources of the same class by leveraging differences in their spatial cues.
In addition, $K$ can range from $0$ to $K_\textrm{max}$, where $0$ means no target sound event in the mixture. 

\subsection{Evaluation method and metric}
\label{sec:metrics}
As the task setting changes, the class-aware signal-to-distortion ratio improvement (CA-SDRi) metric in DCASE25T4 becomes invalid due to the ambiguity introduced by duplicated labels.
We adopt a new metric, class-aware permutation-invariant SDRi (CAPI-SDRi) \cite{dcase2026t4_baseline}, with the core idea is to use the permutation invariant objective to solve the duplicated label ambiguity.

Since the labels in one mixture can be duplicated, 
we denote the unique labels in $C$ and $\hat{C}$ by $\mathcal{C} = \textrm{set}(C)$ and $\hat{\mathcal{C}} = \textrm{set}(\hat{C})$, respectively.
Let $\bar{c}$ be a label in either or both $\mathcal{C}$ or $\hat{\mathcal{C}}$, the collections of estimated and reference waveforms associated with this label are defined as
\begin{equation}
\vspace{-0.1cm}
\begin{aligned}
	S^{\bar{c}} = (\bm{s}_{c_k} \in S \mid c_k = \bar{c}) = (\bm{s}^{\bar{c}}_1,\dots, \bm{s}^{\bar{c}}_{|S^{\bar{c}}|}),\\
	\hat{S}^{\bar{c}} = (\hat{\bm{s}}_{\hat{c}_k} \in \hat{S} \mid \hat{c}_k = \bar{c}) = (\hat{\bm{s}}^{\bar{c}}_1,\dots, \hat{\bm{s}}^{\bar{c}}_{|\hat{S}^{\bar{c}}|}),
\end{aligned}
\end{equation}
where $\bm{s}^{\bar{c}}_i$ and $\hat{\bm{s}}^{\bar{c}}_i$ denote the $i$-th elements of $S^{\bar{c}}$ and $\hat{S}^{\bar{c}}$, respectively, and, $|\cdot|$ indicates the size of the collection.
The counts of true positives (TP), false negatives (FN), and false positives (FP) corresponding to the label $\bar{c}$ can be calculated as
\begin{equation}
\vspace{-0.1cm}
\begin{aligned}
N_{\mathrm{TP}}^{\bar{c}} = \min\bigl(|S^{\bar{c}}|, &|\hat{S}^{\bar{c}}|\bigr), \quad
N_{\mathrm{FN}}^{\bar{c}} = \bigl(|S^{\bar{c}}| - |\hat{S}^{\bar{c}}|\bigr)_+,\\
N_{\mathrm{FP}}^{\bar{c}} &= \bigl(|\hat{S}^{\bar{c}}| - |S^{\bar{c}}|\bigr)_+,
\end{aligned}
\end{equation}
where $(x)_+ = \max(0,x)$. Note that either $N_{\mathrm{FN}}^{\bar{c}}$, $N_{\mathrm{FP}}^{\bar{c}}$, or both are zero for each $\bar{c}$. 
The total number of true and false predictions is
\begin{equation}
\vspace{-0.1cm}
N^{\bar{c}} = N_{\mathrm{TP}}^{\bar{c}} + N_{\mathrm{FN}}^{\bar{c}} + N_{\mathrm{FP}}^{\bar{c}} 
    = \max\bigl(|S^{\bar{c}}|, |\hat{S}^{\bar{c}}|\bigr).
\end{equation}

For true predictions, the waveform metric is calculated. Specifically, the SDRi is evaluated on $N_{\mathrm{TP}}^{\bar{c}}$ sources selected from $S^{\bar{c}}$ and $N_{\mathrm{TP}}^{\bar{c}}$ from $\hat{S}^{\bar{c}}$, with the selection performed using a permutation-invariant objective to maximize the average metric.
The remaining $N_{\mathrm{FN}}^{\bar{c}}$ sources in $S^{\bar{c}}$ and $N_{\mathrm{FP}}^{\bar{c}}$ in $\hat{S}^{\bar{c}}$ are considered false predictions and penalized.
The metric component for the label $\bar{c}$ is defined as
\begin{equation}\label{eq:metric_comp}
\vspace{-0.1cm}
\begin{split}
P^{\bar{c}}
	=
		N^{\bar{c}}_\textrm{FN}\mathcal{P}_\textrm{FN} +
		N^{\bar{c}}_\textrm{FP}\mathcal{P}_\textrm{FP} + 
		\max_{\substack{
			\sigma \in \mathfrak{S}_{|\hat{S}^{\bar{c}}|, N^{\bar{c}}_\textrm{TP}}\\
			\pi \in \mathfrak{C}_{|S^{\bar{c}}|, N^{\bar{c}}_\textrm{TP}}
			}} 
		\sum_{i = 1}^{N^{\bar{c}}_\textrm{TP}} \textrm{SDRi}(\hat{\bm{s}}_{\sigma (i)}^{\bar{c}}, \bm{s}_{\pi (i)}^{\bar{c}}, \bm{y}) ,
\end{split}
\end{equation}
where
\begin{equation}
	\textrm{SDRi}(\hat{\bm{s}}, \bm{s}, \bm{y})
	= \textrm{SDR}(\hat{\bm{s}}, \bm{s}) - \textrm{SDR}(\bm{y} , \bm{s}).
\end{equation}
$\bm{y}$ is the waveform at the reference channel of $\bm{Y}$.
$\mathfrak{S}_{K,L}$ denotes the set of all permutations, and $\mathfrak{C}_{K,L}$ denotes the set of all combinations (i.e., without permutation) of $L$ distinct indices chosen from $\{1, \dots, K\}$.
The penalty values $\mathcal{P}_\textrm{FN}$ and $\mathcal{P}_\textrm{FP}$ are both set to $0$ following \cite{dcase2025t4}.
The mixture-level metric is the average of $P^{\bar{c}}$ as
\begin{equation} \label{eq:metric}
	\textrm{CAPI-SDRi}(\hat{S}, S, \hat{C}, C, \bm{y}) =
	\frac{1}{\sum_{\bar{c} \in \mathcal{C} \cup \hat{\mathcal{C}}} N^{\bar{c}}} 
	\sum_{\bar{c} \in \mathcal{C} \cup \hat{\mathcal{C}}} P^{\bar{c}}.
\end{equation}
The ranking metric is finally averaging the CAPI-SDRi across all the mixture.

For zero-target mixtures, there are no true positives (TP) or false negatives (FN). If false positives (FP) occur, the metric is computed normally using (\ref{eq:metric}). When there are no FP, i.e., a correct silence prediction, (\ref{eq:metric}) is undefined since $\mathcal{C} \cup \hat{\mathcal{C}} = \varnothing$. In such cases, the mixture is excluded from the final averaging. Hence, correctly predicting silence contributes nothing, whereas false predictions are penalized.

It is worth noting that CAPI-SDRi is identical to CA-SDRi in DCASE25T4 when all source labels in the mixture are distinct, while also supporting mixtures containing multiple sources of the same class.
In addition to the main ranking metric, submissions are evaluated using other metrics—such as CASA, PESQ, STOI, and PEAQ—to provide complementary perspectives on performance. 

\begin{table*}[t]
  \centering
  \caption{\footnotesize The amount of isolated target sound event data in the development set used to synthesize the dataset for DCASE 2026 Challenge Task 4, excluding publicly available datasets. In actual training, these data are combined with publicly available datasets~\cite{fsd50k, ears, semhear}; for relatively scarce classes such as Speech, those external data substantially supplement the training material.}
  \label{tab:dataset_summary}

  \setlength{\tabcolsep}{4.2pt}
  \renewcommand{\arraystretch}{0.95}
  \scriptsize

  \resizebox{\textwidth}{!}{%
    \begin{tabular}{ll@{\hspace{6pt}}*{18}{c}}
      \toprule
        & &
        \shortstack[c]{Alarm\\Clock} &
        Blender &
        Buzzer &
        Clapping &
        Cough &
        \shortstack[c]{Cupboard\\OpenClose} &
        Dishes &
        Doorbell &
        \shortstack[c]{Foot\\Steps} &
        \shortstack[c]{Hair\\Dryer} &
        \shortstack[c]{Mechanical\\Fans} &
        \shortstack[c]{Musical\\Keyboard} &
        \shortstack[c]{Percus\\sion} &
        Pour &
        Speech &
        Typing &
        \shortstack[c]{Vacuum\\Cleaner} &
        \shortstack[c]{Bicycle\\Bell} \\
      \midrule
      \multirow{2}{*}{Dev}
        & duration [min]
        & 10.2 & 13.6 & 7.6 & 13.4 & 9.7 & 10.0 & 11.5 & 5.9
        & 24.1 & 39.6 & 40.5 & 13.5 & 21.0 & 10.1 & 5.7 & 17.8 & 39.8 & 4.8 \\
        & \# of samples
        & 78 & 96 & 104 & 119 & 165 & 125 & 90 & 71
        & 158 & 20 & 29 & 108 & 122 & 100 & 77 & 139 & 13 & 89 \\
      \bottomrule
    \end{tabular}%
  }
  \vspace{-15pt}
\end{table*}

\vspace{-5pt}

\section{Dataset for DCASE 2026 Challenge Task 4}
\label{sec:dataset}
\vspace{-5pt}

For DCASE 2026 Challenge Task 4, we newly collected isolated target sound events, first-order Ambisonics room impulse responses (RIRs), and background-noise recordings. We combined them with screened recordings from the dataset released for DCASE 2025 Challenge Task 4 and with selected publicly available data. Using these components, we constructed the dataset for DCASE 2026 Challenge Task 4\footnote{\url{https://doi.org/10.5281/zenodo.19328046}}. The synthesized mixtures follow the revised task setting. They include mixtures with multiple same-class target sources and mixtures without target sound events. The mixtures were generated with \newmodule\footnote{\url{https://github.com/nttcslab/SpAudSyn}}, our spatial-audio synthesis tool developed for this task. The recorded components and the mixture generation procedure are described below.

\subsection{Recorded components}

As described by (\ref{eq:mixture}), each mixture is synthesized from target sound events, optional interference sound events, RIRs, and background noise. The recorded components used to construct these mixtures are summarized as follows.
\begin{itemize}
    \item \textbf{Target sound events}: The target pool covers 18 classes. In this recording, 1053 new isolated target-event recordings were collected across these classes. In addition, 650 recordings from the collection used for the dataset released for DCASE 2025 Challenge Task 4 were re-screened and retained. This resulted in a screened internal pool of 1703 recordings. These recordings were acquired in an anechoic environment with three directional microphones arranged around a source area and one omnidirectional microphone placed above it. Human-produced classes were recorded from 4--8 participants depending on the class, and object-based classes were recorded using 5--15 devices or object sets. For the development set, the screened internal recordings were further combined with curated clips from publicly available datasets such as FSD50K, EARS, and Semantic Hearing. Table~\ref{tab:dataset_summary} summarizes the target-event components used in the development set.
    \item \textbf{Interference sound events}: Sound events from classes distinct from the target events, derived from background set in the dataset of Semantic Hearing \cite{semhear}.
    \item \textbf{Room impulse responses}: In addition to the RIRs carried over from the dataset released for DCASE 2025 Challenge Task 4 and from FOA-MEIR~\cite{foameir}, new first-order Ambisonics RIRs were measured in six rooms with two microphone-array placements per room, corresponding to a center position and a side position. For each placement, a loudspeaker was placed every 20$^{\circ}$ over 18 azimuths on the horizontal plane, at elevations of $0^{\circ}$ and $\pm 20^{\circ}$, and at two source-array distances of 75\,cm and 120 or 150\,cm depending on the room, yielding 1,296 RIRs in total. The RIRs were recorded with a Sennheiser AMBEO VR Mic and converted to B-format.
    \item \textbf{Background noises}: Multi-channel diffuse environmental sounds were recorded with the same Sennheiser AMBEO VR Mic and converted to B-format. The recordings cover seven indoor and outdoor locations, with approximately 15 minutes recorded at each location. During screening, segments containing target sound events were removed from the noise recordings.
\end{itemize}

\subsection{Mixture generation with \newmodule}

The dataset released for DCASE 2025 Challenge Task 4 was synthesized using a modified version of SpatialScaper \cite{spatialscaper}. SpatialScaper was originally developed to simulate and augment data for Sound Event Localization and Detection (SELD), and its original design does not explicitly provide dry reference signals required for source separation. To support the present task, we developed \newmodule for the dataset construction. The functional interface is inspired by SpatialScaper, but \newmodule provides flexible return options for labels, dry sources, metadata, and RIRs. Each sound scene can be fully parameterized and stored as a JSON file, which enables exact mixture reconstruction and reproducible experiments. The implementation is also designed for on-the-fly generation during training.

In the present dataset, each mixture is synthesized by convolving target and interference sources with first-order Ambisonics RIRs and by adding background noise as in (\ref{eq:mixture}). The mixture duration is 10\,s, and all distributed mixtures are provided at 32\,kHz/16-bit. Each mixture contains zero to three target sound events and zero to two interference sound events. The SNR of each target sound event is uniformly sampled between 5 and 20\,dB relative to the background-noise level, whereas interference events are mixed at 0 to 15\,dB. The maximum number of overlapping events is three.

To reflect the revised task setting, part of mixtures contain multiple same-class target sources. 
When such mixtures are generated, the corresponding source directions are separated by at least $60^{\circ}$. 
In other words, even within the same class, spatial information can serve as a clue for distinguishing between them.

A subset of the mixtures contains zero target sound events. These mixtures include only background noise and, optionally, interference sounds. This subset is intended to represent event-absent intervals in long recordings from real applications. For such mixtures, the correct system behavior is to output no detected target events and no separated signals.

\subsection{Development dataset}

The development set is divided into three subsets: training, validation, and test.
The training split provides isolated sources, RIRs, background noises, and interference sounds, and mixtures are generated on the fly during training. The validation split is distributed with fixed metadata so that the same mixtures can be reconstructed for model selection. The test split consists of fixed synthesized mixtures for local evaluation before submission. The validation split contains 1,800 mixtures and the test split contains 1,512 mixtures. In both splits, 16.7\% of mixtures contain no target sound events, 16.7\% contain one target sound event, 33.3\% contain two target sound events, and 33.3\% contain three target sound events. Within the two-target and three-target subsets, 50.0\% of mixtures contain multiple same-class target sources.

\subsection{Evaluation dataset}
The evaluation dataset comprises 2,567 soundscapes, of which the first 1,512 soundscapes are used for ranking in the DCASE 2026 Challenge Task 4, while the remaining soundscapes are used for task analysis. Each soundscape is a 4-channel audio mixture in ambisonic FOA B-format (WYZX), 10 seconds long and is sampled at 32 kHz/16-bit.

The dataset includes synthesized soundscapes and real-world recorded soundscapes.
The synthesized soundscapes are generated using settings similar to those used for the validation set in the development dataset, except that all components--including individual sound events, background noise, and FOA room impulse responses (RIRs)--are newly recorded.
In the real-world subset, audio mixtures are recorded directly using an FOA microphone, the same microphone used to record the RIRs.

For a fair ranking, all the components for constructing the evaluation dataset were newly recorded.
Further details are omitted here because the evaluation split was reserved for challenge ranking.

\section{Baseline system}
\label{sec:baseline}
\begin{figure}[t]
    \centering
    \includegraphics[width=\linewidth]{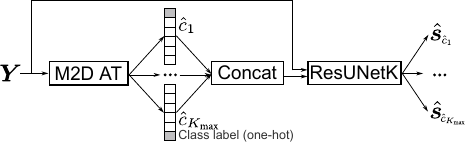}
    \caption{Overview of the baseline system.}
    \label{fig:baseline_system}
\end{figure}

Fig.~\ref{fig:baseline_system} illustrates the overview of the baseline system.
The baseline follows a two-stage pipeline proposed in~\cite{dcase2026t4_baseline}: audio tagging (AT) for sound event classification, followed by label-queried source separation (LQSS).
In the AT stage, a model based on the Masked Modeling Duo (M2D)~\cite{m2d} framework identifies the target sound event classes present in the mixture.
Two AT variants are provided: M2DAT\_1c, which operates on single-channel (omnidirectional) input, and M2DAT\_4c, which operates on 4-channel first-order Ambisonics input.
The latter exploits spatial information at the tagging stage.
In the separation stage, the detected labels are used to query a ResUNetK model, which separates the corresponding source signals from the mixture.

Compared with the DCASE25T4 baseline~\cite{dcase2025t4}, two key changes have been made.
First, the AT model outputs multiple one-hot vectors rather than a single multi-hot vector, enabling the detection of multiple sources belonging to the same class.
Second, a 4-channel AT variant (M2DAT\_4c) is newly introduced.
The ResUNetK architecture itself remains unchanged.
The AT and separation stages are trained independently.

The ResUNetK model is trained using oracle detection labels as input, and its loss function is based on a class-aware permutation-invariant SDR (CA-PI-SDR)~\cite{dcase2026t4_baseline}.
Since oracle labels are used during training, false positives and false negatives do not occur. Therefore, this loss function can also be simply interpreted as a permutation-invariant SDR that takes into account overlapping sources of the same class.

\section{Experimental results}
\label{sec:experiments}

Table~\ref{tab:baseline_results} shows the CAPI-SDRi and classification accuracy of the two baseline systems on the test subset of the development set.
We report two types of accuracy: mixture-level accuracy (Acc.\ (mix)), which measures whether all source labels in a clip are correctly predicted, averaged over all clips; and source-level accuracy (Acc.\ (src)), which measures per-source detection accuracy averaged over all sources.
M2DAT\_4c consistently outperforms M2DAT\_1c across all metrics, indicating that exploiting spatial information in the AT stage is beneficial for both detection and separation.

\begin{table}[t]
    \centering    
    \caption{Baseline results on the test subset of the development set. Acc.\ (mix) denotes mixture-level accuracy and Acc.\ (src) denotes source-level accuracy.}
    \label{tab:baseline_results}
    \scalebox{0.85}[0.85]{
    \begin{tabular}{lccc} \toprule
    System & CAPI-SDRi $\uparrow$ & Acc.\ (mix) $\uparrow$ & Acc.\ (src) $\uparrow$ \\ \midrule
    M2DAT\_1c + ResUNetK & $8.17$ & $57.14$ & $67.15$ \\
    M2DAT\_4c + ResUNetK & $8.49$ & $60.71$ & $70.39$ \\ \bottomrule
    \end{tabular}
    }
\end{table}

\section{Conclusion}
This paper describes the task setting of the DCASE 2026 Challenge Task 4 on Spatial Semantic Segmentation of Sound Scenes. Compared with its first version in \dcaseTwentyFiveTask, \dcaseTwentySixSTask incorporates more realistic conditions, with mixtures having multiple same-class sources or no target sound events. The evaluation metric is also updated accordingly, employing a permutation-invariant objective to address the ambiguity introduced by same-class sources. We construct a new dataset for \dcaseTwentySixSTask based on the \dcaseTwentyFiveTask dataset and newly recorded data, including isolated target sound sources, interference sources, room impulse responses (RIRs), and background noise.


\bibliographystyle{IEEEtran}
\bibliography{refs}

@inproceedings{dcase2025t4,
  title={Description and Discussion on DCASE 2025 Challenge Task 4: Spatial Semantic Segmentation of Sound Scenes},
  author={Yasuda, Masahiro and Nguyen, Binh Thien and Harada, Noboru and Serizel, Romain and Mishra, Mayank and Delcroix, Marc and Araki, Shoko and Takeuchi, Daiki and Niizumi, Daisuke and Ohishi, Yasunori and others},
  booktitle={Workshop on Detection and Classification of Acoustic Scenes and Events (DCASE 2025)},
  year={2025}
}

@inproceedings{dcase2026t4_baseline,
  title={CLASS-AWARE PERMUTATION-INVARIANT SIGNAL-TO-DISTORTION RATIO FOR SEMANTIC SEGMENTATION OF SOUND SCENE WITH SAME-CLASS SOURCES},
  author={Nguyen, Binh Thien and Yasuda, Masahiro and Takeuchi, Daiki and Niizumi, Daisuke and Harada, Noboru},
  booktitle={2026 IEEE International Conference on Acoustics, Speech and Signal Processing (ICASSP)},
  year={2026}
}

@inproceedings{ears,
  title     = {EARS: An Anechoic Fullband Speech Dataset Benchmarked for Speech Enhancement and Dereverberation},
  author    = {Julius Richter and Yi-Chiao Wu and Steven Krenn and Simon Welker and Bunlong Lay and Shinji Watanabe and Alexander Richard and Timo Gerkmann},
  year      = {2024},
  booktitle = {Interspeech 2024},
  pages     = {4873--4877},
  doi       = {10.21437/Interspeech.2024-153},
  issn      = {2958-1796},
}

@INPROCEEDINGS{pesq,
  author={Rix, A.W. and Beerends, J.G. and Hollier, M.P. and Hekstra, A.P.},
  booktitle={2001 IEEE International Conference on Acoustics, Speech, and Signal Processing. Proceedings (Cat. No.01CH37221)}, 
  title={Perceptual evaluation of speech quality (PESQ)-a new method for speech quality assessment of telephone networks and codecs}, 
  year={2001},
  volume={2},
  number={},
  pages={749-752 vol.2},
  doi={10.1109/ICASSP.2001.941023}}

@INPROCEEDINGS{stoi,
  author={Taal, Cees H. and Hendriks, Richard C. and Heusdens, Richard and Jensen, Jesper},
  booktitle={2010 IEEE International Conference on Acoustics, Speech and Signal Processing}, 
  title={A short-time objective intelligibility measure for time-frequency weighted noisy speech}, 
  year={2010},
  volume={},
  number={},
  pages={4214-4217},
  doi={10.1109/ICASSP.2010.5495701}}

@article{m2d,
  title={Masked modeling duo: Towards a universal audio pre-training framework},
  author={Niizumi, Daisuke and Takeuchi, Daiki and Ohishi, Yasunori and Harada, Noboru and Kashino, Kunio},
  journal={IEEE/ACM Trans. on Audio, Speech, and Lang. Process.},
  year={2024},
  publisher={IEEE}
}

@inproceedings{semhear,
  title={Semantic hearing: Programming acoustic scenes with binaural hearables},
  author={Veluri, Bandhav and Itani, Malek and Chan, Justin and Yoshioka, Takuya and Gollakota, Shyamnath},
  booktitle={Proceedings of the 36th Annual ACM Symposium on User Interface Software and Technology},
  pages={1--15},
  year={2023}
}

@article{fsd50k,
  title={{FSD50K}: an open dataset of human-labeled sound events},
  author={Fonseca, Eduardo and Favory, Xavier and Pons, Jordi and Font, Frederic and Serra, Xavier},
  journal={IEEE/ACM Transactions on Audio, Speech, and Language Processing},
  volume={30},
  pages={829--852},
  year={2021},
  publisher={IEEE}
}

@inproceedings{foameir,
  title={Echo-aware adaptation of sound event localization and detection in unknown environments},
  author={Yasuda, Masahiro and Ohishi, Yasunori and Saito, Shoichiro},
  booktitle={IEEE Intl. Conf. on Acoust., Speech \& Sig. Proc. (ICASSP)},
  pages={226--230},
  year={2022}
}

@inproceedings{spatialscaper,
  title={Spatial scaper: a library to simulate and augment soundscapes for sound event localization and detection in realistic rooms},
  author={Roman, Iran R and Ick, Christopher and Ding, Sivan and Roman, Adrian S and McFee, Brian and Bello, Juan P},
  booktitle={IEEE Intl. Conf. on Acoust., Speech \& Sig. Proc. (ICASSP)},
  pages={1221--1225},
  year={2024}
}

\end{document}